 \documentclass[a4paper]{article}
 \usepackage{geometry}
  \usepackage{color}
 \usepackage{amsmath}
 \newcommand{\bv}{\begin{vmatrix}}
 	\newcommand{\ev}{\end{vmatrix}}

 \newcommand{\bea}{\begin{eqnarray*}}
 	\newcommand{\eea}{\end{eqnarray*}}
 \newcommand{\bean}{\begin{eqnarray}}
 	\newcommand{\eean}{\end{eqnarray}}

 \newcommand{\grad}{\nabla}

 \date{ }
 \title{Higher-dimensional charged black holes cannot be over-charged by gedanken experiments}
 \author{Boxuan Ge\footnote{boxuange@mail.bnu.edu.cn},  Yuyu Mo\footnote{yymo@mail.bnu.edu.cn},  Suting Zhao\footnote{stzhao@mail.bnu.edu.cn}, 
 	\\
 	Department of Physics, Beijing Normal University,
 	Beijing 100875, China
 	\\
 	 Jieping Zheng\footnote{jpzheng@mail.bnu.edu.cn}\\
 	Department of Physics and Astronomy, Shanghai Jiao Tong University,
 	Shanghai 200240, China}

\begin{document}
 \maketitle
 \begin{abstract}
 We reconsider over-charging the higher-dimensional nearly extremal charged black holes using the new version of gedanken experiment proposed recently by Sorce and Wald.  As a result, we find that cosmic censorship conjecture associated with such black holes is restored by taking into account the second-order correction, albeit violated by the first-order perturbation. Namely, the higher-dimensional nearly extremal charged black holes cannot be over-charged.
 \end{abstract}
 	\section{Introduction}
 	To ensure the predictability of general relativity as a classical theory,  long time ago Penrose proposed a conjecture, dubbed as weak cosmic censorship(WCC), which states that all singularities caused by gravitational collapse must be shielded by the black holes horizon such that it does not affect the distant observers.
 	
 	Although there is still no general proof for WCC, many efforts have been taken for decades to test it\cite{Wald74}.  Among others, Wald suggested a gedanken experiment to see the possibility of formation of naked singularity by over-charging or over-spinning a black hole with particle matters$\cite{Wald741}$. The result shows that no violation of WCC can occur when one tries to destroy an extremal Kerr-Newman black hole in such a way. However, later on Hubeny found that it is possible for one to violate WCC by over-charging a nearly extremal charged black hole\cite{Hubeny}. The follow-up works further show that the Hubeny type violation of WCC is actually universal to a general nearly extremal Kerr-Newman black hole\cite{v1,v2,app2,v3,v4}. It was later recognized that it is pivotal to carefully incorporate the self-force and finite-size effects of test particle matters before claiming the true violation of WCC in all the Hubeny scenarios, because both of these effects enter at higher order of particle's energy, charge and angular moment while the aforementioned analyses are performed only at linear order\cite{sf1,sf2,sf3}.
	 	
 	To solve this issue and let the dust settle completely, Sorce and Wald has recently invented a new version of gedanken experiment without relying on the test particle assumption or the explicit analyses of trajectories of particle matters\cite{Wald17}. Instead, they apply the Iyer-Wald formalism to generic matter perturbation on top of the black hole in consideration and obtain two perturbation inequalities by imposing the null energy condition on the involved matter, where the first order inequality boils down into the condition for the black hole to capture the particle matter in the old version of gedanken experiment when one takes the particle matter as the limiting case of the generic matter while the second order inequality is demonstrated to encode the self-force and finite-size effects in an elegant way. In particular, they reexamine the Hubeny scenarios by this new gedanken experiment and find that a nearly extremal Kerr-Newman black hole cannot be over-charged or over-spun when the second order perturbation inequality is taken into account.
		
    Most recently, this set-up has also been used to explore the possibility of over-spinning the five-dimensional Myers-Perry black holes in \cite{jincheng}, and it is shown that although the Hubeny type violation of WCC can occur at the linear order, the second order perturbation inequality prevents a five-dimensional nearly extremal Myers-Perry black hole from being over-spun. Therefore it is natural for us to ask whether the second order perturbation inequality enables WCC to be restored from all the Hubeny type violations. In particular, it has been shown recently by Revelar and Vega that it is possible to lead to the Hubeny type violation of WCC by  over-charging $n>4$ dimensional nearly extremal charged black hole in the old version of gedanken experiment, albeit only in a small region of parameter space\cite{Karl}. The purpose of this paper is to reconsider this Hubeby scenario by the new version of gedanken experiment and investigate whether such a violation can still occur when the second order correction is incorporated.
 	
 	The rest of this paper is structured as follows. In Section \ref{formalism}, we briefly review the Iyer-Wald formalism for a general diffeomorphism covariant theory and the corresponding variational  identity. In the subsequent section, we restrict ourselves on the higher-dimensional Einstein-Maxwell theory, where the explicit expressions for some relevant quantities are presented and the static charged black hole solutions are introduced. In Section \ref{gedanken}, we present the set-up for the new version of gedanken experiment on top of the non-extremal charged black holes, and derive the first and second order inequalities for our perturbation. In Section \ref{destory}, we examine the Hubeny scenario by conducting the gedanken experiment on top of nearly extremal charged black holes and verify that for a higher-dimensional nearly extremal charged black hole, no violation of Hubeny type can occur when the second order correction is considered. In final section, we conclude our paper with some discussions.

 	\section{Iyer-Wald Formalism and Variational Identities}\label{formalism}
 	We consider a diffeomorphism covariant theory on an $n$-dimensional oriented manifold $\mathcal{M}$, where the Lagrangian $n$-form $\mathbf{L}=L\mathbf{\epsilon}$ is supposed to be constructed locally out of the metric $g_{ab}$, other matter fields $\psi$, as well as the symmetrized covariant derivatives of the corresponding Riemann tensor $R_{abcd}$ and $\psi$, with $\epsilon$ the volume element compatible with the metric on the manifold $\mathcal{M}$\cite{Wald94}. We use $\phi=(g_{ab},\psi)$ to denote all dynamical fields and perform a variation of $\mathbf{L}$, which leads to
 	\begin{eqnarray}
 		\delta\mathbf{L}=\mathbf{E}\delta\phi + d\mathbf{\Theta}(\phi,\delta\phi),\label{1}
 	\end{eqnarray}
 	where
  $\mathbf{E}=0$ correspond to the equations of motion of the theory, and $\mathbf{\Theta}$  is called the symplectic potential $(n-1)$-form. The symplectic current $(n-1)$-form $\omega$ is then defined by
 	\begin{eqnarray}
 		\mathbf{\omega}(\phi,\delta_{1}\phi,\delta_{2}\phi)=\delta_{1}\mathbf{\Theta}(\phi,\delta_{2}\phi)-\delta_{2}\mathbf{\Theta}(\phi,\delta_{1}\phi),\label{2}
 	\end{eqnarray}
 	where $\delta_{1}$ and $\delta_{2}$ denote the variations with respect to different parameters.
 	
 	The Noether current $(n-1)$-form $\mathbf{J}_{\chi}$ associated with an arbitrary smooth vector field $\chi^a$ is defined as
 	\begin{eqnarray}
 		\mathbf{J}_{\chi}=\mathbf{\Theta}(\phi,\mathcal{L}_{\chi}\phi)-\chi\cdot\mathbf{L},\label{3}
 	\end{eqnarray}
 	where we replace $\delta$ by $\mathcal{L}_{\chi}$ in the expression of $\mathbf{\Theta}$ and the `dot' represents the contraction of $\chi^a$ into the first index of $\mathbf{L}$. A simple calculation gives
 	\begin{eqnarray}
 		d\mathbf{J}_{\chi}=-\mathbf{E}\mathcal{L}_{\chi}\phi,\label{4}
 	\end{eqnarray}
 	which indicates $d\mathbf{J}_{\chi}=0$ when the equations of motion are satisfied. On the other hand, as shown in \cite{Wald95}, the Noether current $(n-1)$-form can also be expressed in the following form
 	\begin{eqnarray}
 		\mathbf{J}_{\chi}=\mathbf{C}_{\chi}+d\mathbf{Q}_{\chi}.\label{5}
 	\end{eqnarray}
 Here $\mathbf{Q}_{\chi}$ is the so-called Noether charge associated with $\chi^a$ and $\mathbf{C}_{\chi}=\chi \cdot \mathbf{C}$ are interpreted as the corresponding constraints of the theory, which vanish when the equations of motion are satisfied. By comparing the variations of equations \eqref{3} and \eqref{5} with $\chi^a$ fixed, we obtain the first variational identity
 	\begin{eqnarray}
 		d[\delta \mathbf{Q}_{\chi}-\chi\cdot\mathbf{\Theta}(\phi,\delta\phi)]=\omega(\phi,\delta\phi,\mathcal{L}_{\chi}\phi)-\chi\cdot\mathbf{E}\delta\phi-\delta\mathbf{C}_{\chi}.\label{6}
 	\end{eqnarray}
	The variation of this first variational identity further gives rise to the second variational identity
	\begin{eqnarray}
 		d[\delta^2 \mathbf{Q}_{\chi}-\chi\cdot\delta\mathbf{\Theta}(\phi,\delta\phi)]=\omega(\phi,\delta\phi,\mathcal{L}_{\chi}\delta\phi)-\chi\cdot\delta\mathbf{E}\delta\phi-\delta^2\mathbf{C}_{\chi},\label{7}
 	\end{eqnarray}	
	where we have used the equations of motion $\mathbf{E}=0$ and assumed that $\chi^a$ is a symmetry of $\phi$, i.e., $\mathcal{L}_{\chi}\phi=0$.
 	
 \section{Einstein-Maxwell Theory and Higher-dimensional Charged Black Holes}\label{einstein}
 	For our purpose, we now consider the Einstein-Maxwell theory in $n$-dimensional spacetime. The corresponding Lagrangian $n$-form reads
 	\begin{eqnarray}
 		\mathbf{L}=\frac{1}{16\pi}(R-F^{ab}F_{ab})\mathbf{\epsilon},\label{8}
 	\end{eqnarray}
 	where we have set the Newton constant $G=1$.
 	The variation of the Lagrangian $n$-form gives
 	\begin{eqnarray}
 		\mathbf{E}(\phi)\delta\phi=-\mathbf{\epsilon}(\frac{1}{2} T^{ab}
 		\delta g_{ab}+j^a\delta A_{a}),\label{9}
 	\end{eqnarray}
 	where
 	\begin{eqnarray}
 		8\pi T_{ab}\equiv G_{ab}-8\pi T_{ab}^{EM}, \quad j^a=\frac{1}{4\pi} \grad_{b} F^{ab}\label{10}
		\end{eqnarray}
		with
		\begin{eqnarray}
 		T_{ab}^{EM}=\frac{1}{4\pi}(F_{ac}F_{b}{}^{c}-\frac{1}{4}g_{ab}F^{cd}F_{cd}).\label{11}
 	\end{eqnarray}
 	On the other hand, the symplectic potential $(n-1)$-form is given by
 	\begin{eqnarray}
 		\mathbf{\Theta}(\phi,\delta\phi)=\mathbf{\Theta}^{GR}(\phi,\delta\phi)+\mathbf{\Theta}^{EM}(\phi,\delta\phi)\label{12}
 	\end{eqnarray}
 	with
 	\begin{eqnarray}
 		&&\Theta^{GR}_{a_{2}...a_{n}} (\phi,\delta\phi)=\frac{1}{16\pi}\epsilon_{d a_{2}...a_{n}}g^{de} g^{fg}(\grad_{g}\delta g_{ef}-\grad_{e}\delta g_{fg}),\label{13}\\
 		&&\Theta^{EM}_{a_{2}...a_{n}}(\phi,\delta\phi)=-\frac{1}{4\pi}\epsilon_{d a_{2}...a_{n}}F^{de}\delta A_{e}.\label{14}
 	\end{eqnarray}
 Whence the symplectic current can be obtained as
 	\begin{eqnarray}
 		\mathbf{\omega}(\phi;\delta_{1}\phi,\delta_{2}\phi)=\mathbf{\omega}^{GR}(\phi;\delta_{1}\phi,\delta_{2}\phi)+\mathbf{\omega}^{EM}(\phi;\delta_{1}\phi,\delta_{2}\phi),\label{15}
 	\end{eqnarray}
 	where
 	\begin{eqnarray}
 		&&\omega_{a_{2}...a_{n}}^{GR}=\frac{1}{16\pi}\epsilon_{d a_{2}...a_{n}} w^{d},\label{16}\\
 		&&\omega_{a_{2}...a_{n}}^{EM}=\frac{1}{4\pi}[\delta_{2}(\epsilon_{d a_{2}...a_{n}} F^{de}) \delta_{1} A_{e}-\delta_{1}(\epsilon_{d a_{2}...a_{n}} F^{de}) \delta_{2} A_{e}]\label{17}
 	\end{eqnarray}
 	with
 	\begin{eqnarray}
 		&& w^{a}=P^{abcdef}(\delta_{2} g_{bc} \grad_{d}\delta_{1}g_{ef}-\delta_{1} g_{bc} \grad_{d}\delta_{2}g_{ef}),\label{18}\\
 		&& P^{abcdef}=g^{ae} g^{fb} g^{cd}-\frac{1}{2}g^{ad} g^{be} g^{fc} - \frac{1}{2}g^{ab} g^{cd} g^{ef} - \frac{1}{2}g^{bc} g^{ae} g^{fd} + \frac{1}{2}g^{bc} g^{ad} g^{ef}.\label{19}
 	\end{eqnarray}
 	By using $\mathcal{L}_{\chi} g_{ab} = \grad_{a}\chi_{b} +\grad_{b}\chi_{a}$ and $\grad_{a}A_{b}=F_{ab}+\grad_{b}A_{a}$, one can calculate out the Noether current $(n-1)$-form $\mathbf{J}_{\chi}$  in a straightforward way as
	  	\begin{eqnarray}
 		(J_{\chi})_{a_{2}...a_{n}}=(J^{GR}_{\chi})_{a_{2}...a_{n}} + (J^{EM}_{\chi})_{a_{2}...a_{n}},\label{20}
 	\end{eqnarray}
 	where
 	\begin{eqnarray}
 		&&(J^{GR}_{\chi})_{a_{2}...a_{n}}=\frac{1}{8\pi}\epsilon_{e a_{2}...a_{n}} \grad_{f}(\grad^{[f} \chi^{e]}) + \epsilon_{e a_{2}...a_{n}} T_{f}{}^{e} \chi^{f},\label{21}\\
 		&&(J^{EM}_{\chi})_{a_{2}...a_{n}}=\frac{1}{4\pi}\epsilon_{e a_{2}...a_{n}}\grad_{g}(F^{ge} A_{f} \chi^{f}) + \epsilon_{e a_{2}...a_{n}} A_{f} j^{e} \chi^{f},\label{22}
 	\end{eqnarray}
 	Then it is not hard for us to identify the Noether charge
 	\begin{eqnarray}
 		(Q_{\chi})_{a_{3}...a_{n}}=(Q_{\chi}^{GR})_{a_{3}...a_{n}}+(Q_{\chi}^{EM})_{a_{3}...a_{n}}  \label{23}
 	\end{eqnarray}
 	with
 	\begin{eqnarray}
 		&&(Q_{\chi}^{GR})_{a_{3}...a_{n}}=-\frac{1}{16\pi}\epsilon_{d e a_{3}...a_{n}}\grad^{d}\chi^{e},\label{24}\\
 		&&(Q_{\chi}^{EM})_{a_{3}...a_{n}}=-\frac{1}{8\pi}\epsilon_{d e a_{3}...a_{n}} F^{d e} A_{f} \chi^{f},\label{25}
 	\end{eqnarray}
 	as well as the constraint
 	\begin{eqnarray}
 		(C_{f})_{a_{2}...a_{n}}=\epsilon_{e a_{2}...a_{n}}(T_{f}{}^{e} + A_{f} j^{e}).\label{26}
 	\end{eqnarray}
 	
 	We now restrict on the static spherically symmetric charged black hole solution to the Einstein-Maxwell theory in $n$-dimensional spacetime, which is also called the charged Schwarzschild-Tangherlini black hole. The corresponding line element and electric potential read
 	\begin{eqnarray}
 		d s^{2} = -f(r) dt^{2} + f(r)^{-1} dr^{2} + r^{2} d\Omega_{n-2}^{2}, \quad A_{a}=-\frac{4\pi Q}{(n-3)\Omega r^{n-3}}(dt)_{a}, \label{27}
 	\end{eqnarray}
 	where
 	\begin{eqnarray}
 		f(r) = 1 - \frac{\mu}{r^{n-3}} + \frac{\nu^2}{r^{2(n-3)}},\label{28}
 	\end{eqnarray}
 	$\Omega=2\pi^{(n-1)/2}/\Gamma[(n-1)/2]$ is  the volume of the unit $(n-2)$-sphere
 	\begin{eqnarray}
 		d\Omega_{n-2}^{2}=d\theta_{1}^{2}+\sin^2\theta_{1}d\theta_{2}^{2}+...+\sin^{2}\theta_{1}\cdots\sin^{2}\theta_{n-3}d\theta_{n-2}^{2},\label{29}
 	\end{eqnarray}
 and $Q$ is the electric charge of the black hole. Furthermore the parameters $\mu$ and $\nu$ are related to the ADM mass $M$ and the charge $Q$ of the black hole as
 	\begin{eqnarray}
 		\mu=\frac{16\pi M}{(n-2)\Omega},\qquad
 		\nu=\frac{4\pi}{\Omega}\sqrt{\frac{2}{(n-2)(n-3)}} Q.\label{30}
 	\end{eqnarray}
	The spacetime singularity is located at $r=0$, which is naked when $\mu^{2}-4\nu^{2}<0$ and benign when $\mu^{2}-4\nu^{2}\geq0$, because in the latter case the singularity is well shielded by the black hole horizons
	\begin{eqnarray}
 		r_{\pm}^{n-3} = \frac{\mu\pm\sqrt{\mu^2-4\nu^2}}{2},\label{31}
 	\end{eqnarray}	
	which are obtained by solving $f(r)=0$ and coincide with each other at the extremality $\mu^{2}-4\nu^{2}=0$. In what follows, we shall focus exclusively on $r_+$, which corresponds to the event horizon with the area
 	\begin{eqnarray}
 		A=\Omega r_{+}^{n-2}.\label{32}
 	\end{eqnarray}
 	Note that our event horizon is a Killing horizon generated by the Killing field $\xi^{a}=(\frac{\partial}{\partial t})^{a}$, so the corresponding horizon potential and surface gravity are given by
 	\begin{eqnarray}
 		\Phi_{H}=-\xi^a A_{a}|_{r_{+}}=\frac{4\pi Q}{(n-3)\Omega r_{+}^{n-3}},\label{33}
 	\end{eqnarray}
 and
 	\begin{eqnarray}
 		\kappa=\frac{1}{2}f'(r)|_{r_{+}}=\frac{n-3}{2}(\frac{\mu}{r_{+}^{n-2}}-\frac{2\nu^{2}}{r_{+}^{2n-5}}).\label{34}
 	\end{eqnarray}

 	\section{Perturbation Inequalities of Gedanken Experiments}\label{gedanken}
 	Following the idea in \cite{Wald17}, we now want to present the set-up for the new version of gedanken experiment and investigate what the allowed perturbation is when one tries to over-charge a nearly-extremal charged black hole. To be more precise, let us consider an one-parameter family of field configurations $\phi(\lambda)$ with $\phi_{0}=\phi(0)$ corresponds to an $n$-dimensional nearly extremal charged Schwarzschild-Tangherlini black hole. For such a non-extremal black hole , the horizon is of bifurcate type. Thus we can choose a hypersurface $\Sigma=\Sigma_1 \cup \mathcal{H}$, such that it starts from the bifurcation surface $B$ and continues up the future horizon through the portion $\mathcal{H}$ till the cross section $S$, then becomes spacelike as $\Sigma_1$ to extend towards the spatial infinity. For simplicity but without loss of generality, we shall assume that all matter sources go into the black hole through a finite portion of $\mathcal{H}$, so that there is no perturbation in the vicinity of $B$ and the perturbed matter sources vanish in the neighborhood of $S$ as well as on $\Sigma_{1}$. This implies the following asymptotical behaviors
	\begin{eqnarray}\label{asym}
	g_{\mu\nu}(\lambda)=\eta_{\mu\nu}+\mathcal{O}(\frac{1}{r^{n-3}}), \quad \partial_\rho g_{\mu\nu}(\lambda)=\mathcal{O}(\frac{1}{r^{n-2}}), \quad A_\mu(\lambda)=\mathcal{O}(\frac{1}{r^{n-3}}), \quad F_{\mu\nu}(\lambda)=\mathcal{O}(\frac{1}{r^{n-2}})
		\end{eqnarray}
	as one approaches on $\Sigma_1$ to the spatial infinity, where the Lorentzian coordinates of a flat metric $\eta_{ab}$ are used. Furthermore, we would like to work with the gauge of Gaussian null coordinates on $\mathcal{H}$, where we especially have
		\begin{eqnarray}
 		\int_{B}\mathbf{Q}_{\xi}^{GR}(\lambda)=\frac{\kappa}{8\pi}A_{B}(\lambda)\label{36}
 	\end{eqnarray}
 	with $A_{B}$ the area of the bifurcate surface\cite{Hollands}.
	
	With the above preparation, we now derive the first-order inequality obeyed by the perturbation at $\lambda=0$. Note that for our choice of Cauchy surface $\Sigma$, the boundaries of $\Sigma$ are located at the bifurcate surface and the spatial infinity, respectively. So the integration of the left-hand side of \eqref{6} on $\Sigma$ reads
 	\begin{eqnarray}
 		\int_{\partial\Sigma}(\delta \mathbf{Q}_{\xi}-\xi \cdot \mathbf{{\Theta}})=\int_{\infty}(\delta \mathbf{Q}_{\xi}-\xi \cdot \mathbf{{\Theta}})-\int_{B}(\delta \mathbf{Q}_{\xi}-\xi \cdot \mathbf{{\Theta}}).\label{37}
 	\end{eqnarray}
	It follows from (\ref{asym}) that there is no electromagnetic contribution to the boundary term at infinity while the corresponding boundary term from the gravitational part is given precisely by the variation of the ADM mass $\delta M$. In addition, note that $\xi^a=0$ at the bifurcate surface. Then it is not hard to see the contribution to boundary term at the bifurcate surface is given by $\frac{\kappa}{8\pi}\delta A_B+\Phi_H\delta Q_B$, which also vanishes according to our assumption that no perturbation occurs in the vicinity of $B$. Thus we end up with
    \begin{eqnarray}
 	    \int_{\partial\Sigma}(\delta \mathbf{Q}_{\xi}-\xi \cdot \mathbf{{\Theta}})=\int_{\infty}(\delta \mathbf{Q}_{\xi}^{GR}-\xi \cdot \mathbf{{\Theta}}^{GR})=\delta M.\label{38}
    \end{eqnarray}
   On the other hand, note that $E(\phi_0)=0$ and $\mathcal{L}_\xi\phi_0=0$, so for the integration of the right-hand side of \eqref{6} on $\Sigma$, the only non-vanishing contribution comes from
    \begin{eqnarray}
 	    -\int_{\Sigma}\delta \mathbf{C}_{\xi}&=&-\delta\int_{\Sigma}\epsilon_{e a_{2}...a_{n}}T_{f}{}^{e}\xi^f - \delta\int_{\Sigma}\epsilon_{e a_{2}...a_{n}}A_{f}\xi^{f}j^{e}\nonumber\\
 	    &=&-\int_{\mathcal{H}}\epsilon_{e a_{2}...a_{n}}\delta T_{f}{}^{e}\xi^f - \int_{\mathcal{H}}\xi^{f}A_{f}\delta(\epsilon_{e a_{2}...a_{n}}j^{e})\nonumber\\
 	    &=&\int_{\mathcal{H}}\widetilde{\epsilon}_{a_{2}...a_{n}}\delta T_{fe}\xi^f k^e + \Phi_{H}\delta Q \ge \Phi_{H}\delta Q.\label{39}
    \end{eqnarray}
     Here $k^a=(\frac{\partial}{\partial u})^a\propto\xi^{a}$ is the future-directed tangent vector field of the affinely parametrized null geodesic generators of the horizon, and $\widetilde{\epsilon}_{a_{2}...a_{n}}$, defined as $\epsilon_{ea_{2}...a_{n}}=-nk_{[e}\widetilde{\epsilon}_{a_{2}...a_{n}]}$, is the volume element on the horizon. In addition, the above inequality is achieved by the null energy condition imposed on the perturbed non-electromagnetic stress-energy tensor on $\mathcal{H}$, i.e., $\delta T_{ab}k^a k^b|_\mathcal{H}\ge 0$. Combining \eqref{38} and \eqref{39}, we obtain the first-order perturbation inequality as
    \begin{eqnarray}
 	    \delta M-\Phi_{H}\delta Q\ge 0.\label{41}
    \end{eqnarray}
     Obviously, if we want to violate $\mu^2-4\nu^2 \ge 0$, the optimal choice is to saturate \eqref{41} by requiring $\delta T_{ab}k^ak^b|_{\mathcal{H}}=0$, namely, the energy flux through the horizon vanishes for the first-order non-electromagnetic perturbation. Then it follows from Raychaudhuri equation
     \begin{eqnarray}
     \frac{d\vartheta(\lambda)}{du}=-\frac{1}{n-2}\vartheta(\lambda)^2-\sigma_{ab}(\lambda)\sigma^{ab}(\lambda)-R_{ab}(\lambda)k^ak^b
     \end{eqnarray}
     that the first-order perturbation of the expansion $\delta\vartheta=0$ if we further require the first-order perturbed horizon coincides with the unperturbed horizon\cite{Gao}\footnote{As shown in \cite{Hollands}, this requirement can always be achieved.}.

    With such an optimal choice, we are in a position to derive the second order inequality for the allowed perturbation. By integrating \eqref{7} on $\Sigma$ and performing a similar analysis to the first-order case, we have
    \begin{eqnarray}
 	    \delta^2 M =\mathcal{E}_{\Sigma}(\phi_{0};\delta\phi)-\int_{\mathcal{H}}\xi\cdot\delta\mathbf{E}\delta\phi-\int_{\mathcal{H}}\delta^2 \mathbf{C}_{\xi}\label{43}
    \end{eqnarray}
    with
    \begin{eqnarray}
        &&(\xi \cdot\delta\mathbf{E}\delta\phi)_{a_{2}...a_{n}}=-\xi^{e}\epsilon_{ea_{2}...a_{n}}(\frac{1}{2}\delta T^{ab}
        \delta g_{ab} + \delta j^a \delta A_{a}),\label{44}\\
        &&(\delta^2\mathbf{C}_{\xi})_{a_{2}...a_{n}}=\delta^2(\epsilon_{e a_{2}...a_{n}}T_{f}{}^{e}\xi^{f}) + \delta^2(\epsilon_{e a_{2}...a_{n}}A_{f} \xi^{f}j^{e})\label{45}
    \end{eqnarray}
    and
    \begin{eqnarray}
        \mathcal{E}_{\Sigma}(\phi_{0};\delta\phi)=\int_{\Sigma}\mathbf{\omega}(\phi_{0};\delta\phi,\mathcal{L}_{\xi}\delta\phi).\label{46}
    \end{eqnarray}

    Since $\xi^{e}$ is tangent to $\mathcal{H}$, the second term of the right side of \eqref{43} makes no contribution when pulled back onto $\mathcal{H}$. On the other hand, we can always impose the condition $\xi^a \delta A_{a}|_{\mathcal{H}}=0$ by a gauge transformation, so \eqref{43} can be rewritten as
    \begin{eqnarray}
        \delta^2 M &=& \mathcal{E}_{\Sigma}(\phi_{0};\delta\phi) -\delta^2\int_{\mathcal{H}}\epsilon_{e a_{2}...a_{n}}\xi^{f} T_{f}{}^{e} - \delta^2\int_{\mathcal{H}}A_{f}\xi^{f}(\epsilon_{e a_{2}...a_{n}}j^{e})\nonumber\\
        &=& \mathcal{E}_{\Sigma}(\phi_{0};\delta\phi) -\int_{\mathcal{H}}\epsilon_{e a_{2}...a_{n}}\xi^{f}\delta^2 T_{f}{}^{e} - \int_{\mathcal{H}}A_{f}\xi^{f}\delta^2(\epsilon_{e a_{2}...a_{n}}j^{e})\nonumber\\
 	    &=&\mathcal{E}_{\Sigma}(\phi_{0};\delta\phi) + \int_{\mathcal{H}}\widetilde{\epsilon}_{a_{2}...a_{n}}\xi^f k^e\delta^2 T_{fe} + \Phi_{H}\delta^{2}Q\nonumber\\
 	    &\ge&\mathcal{E}_{\Sigma_{1}}(\phi_{0};\delta\phi)+\mathcal{E}_{\mathcal{H}}(\phi_{0};\delta\phi)+\Phi_{H}\delta^{2}Q,\label{49}
    \end{eqnarray}
    where we have used $\delta T_{ab}k^ak^b|_{\mathcal{H}}=0$ in the second step, $k_e\delta g^{ed}|_\mathcal{H}=0$ for the Gaussian null coordinates in the third step, and the null energy condition for the second order perturbed non-electromagnetic stress-energy tensor, i.e., $\delta^2 T_{ab} k^a k^b|_\mathcal{H} \ge 0$ in the last step.

  To obtain $\mathcal{E}_{\mathcal{H}}(\phi_{0};\delta\phi)$, we shall make an additional but reasonable assumption that the initial black hole is eventually driven into another charged Schwarzschild-Tangherlini black hole by the linear perturbation. Then after a straightforward but a little bit tedious calculation, one can obtain
    \begin{eqnarray}
        \mathcal{E}_{\mathcal{H}}(\phi_{0};\delta\phi)=\int_{\mathcal{H}}\omega^{GR}(\phi_0;\delta\phi,\mathcal{L}_{\xi}\delta\phi)+\int_{\mathcal{H}}\omega^{EM}(\phi_0;\delta\phi,\mathcal{L}_{\xi}\delta\phi)\label{52}
    \end{eqnarray}
    with the gravitational contribution given by\cite{Hollands}
    \begin{eqnarray}
 	    \int_{\mathcal{H}}\omega^{GR}(\phi_0;\delta\phi,\mathcal{L}_{\xi}\delta\phi)=\frac{1}{4\pi}\int_{\mathcal{H}}\widetilde{\epsilon}_{a_{2}...a_{n}}(\xi^a\grad_{a}u)\delta\sigma_{bc}\delta\sigma^{bc},\label{53}
 	 \end{eqnarray}
 and the electromagnetic contribution given by\cite{Wald17}
 \begin{eqnarray}
  \int_{\mathcal{H}}\omega^{EM}(\phi_0;\delta\phi,\mathcal{L}_{\xi}\delta\phi)=\frac{1}{2\pi}\int_{\mathcal{H}}\widetilde{\epsilon}_{a_{2}...a_{n}}k^d\xi^e\delta F_d{}^{f}\delta F_{ef}.\label{54}
    \end{eqnarray}
   It is obvious to see that both of them are non-negative.

    In addition, to obtain $\mathcal{E}_{\Sigma_{1}}(\phi_{0};\delta\phi)$, we adopt the trick in \cite{Wald17} by considering another one-parameter family of field configurations $\phi^{ST}(\lambda')$, where each field configuration corresponds to an $n$-dimensional charged Schwarzschild-Tangherlini black hole with the mass and charge given by
    \begin{eqnarray}
 	    M^{ST}(\lambda')=M+\lambda'\delta' M,\quad Q^{ST}(\lambda')=Q+\lambda'\delta' Q.\label{55}
    \end{eqnarray}
    Here $\phi^{ST}(0)=\phi_0$, and
     $\delta'$ denotes the derivative with respect to $\lambda'$ at $\lambda'=0$ with $\delta' M=\delta M$ and $\delta' Q=\delta Q$ such that $\delta'\phi^{ST}=\delta \phi$ on $\Sigma_1$. As to this family of field configurations, we apparently have
    \begin{eqnarray}
        \delta'^2 M=\delta'^2 Q_B=\delta'^2 \mathbf{C}_{\xi}=\delta'\mathbf{E}=\mathcal{E}_{\mathcal{H}}(\phi_{0};\delta'\phi^{ST})=0.\label{56}
    \end{eqnarray}
     Thus the integration of \eqref{7} over $\Sigma$ gives rise to
    \begin{eqnarray}
    -\frac{\kappa}{8\pi}\delta'^2 A_{B}^{ST}=\mathcal{E}_{\Sigma_{1}}(\phi_{0};\delta'\phi^{ST})=\mathcal{E}_{\Sigma_1}(\phi_{0};\delta\phi)
    ,\label{60}	
    \end{eqnarray}
    By taking all the above results together, we wind up with the second-order inequality for our perturbation as
    \begin{eqnarray}
 	    \delta^2 M-\Phi_{H}\delta^2Q\ge-\frac{\kappa}{8\pi}\delta'^2 A_{B}^{ST}.\label{62}
    \end{eqnarray}

    \section{Gedanken Experiments to Destroy a Nearly Extremal Charged Black Hole}\label{destory}
    With the previous preparation, we now examine the Hubeny scenario by conducting the new version of gedanken experiment on top of the $n$-dimensional nearly extremal charged black hole and verify that such a black hole can not be over-charged when the second-order correction of the perturbation are taken into consideration. To proceed, we define a function of $\lambda$ as
    \begin{eqnarray}
 	    h(\lambda)=\mu(\lambda)^2-4\nu(\lambda)^2.\label{63}
    \end{eqnarray}
     We write $h(0)=\mu_{0}^2\alpha^2$ with $\mu_{0}=\mu(0)$, $\nu_{0}=\nu(0)$ and $\alpha=\sqrt{1-4\nu_{0}^2/\mu_{0}^2}$. Obviously, $\alpha=0$ corresponds to an extremal black hole, and a positive but small $\alpha$ corresponds to a nearly extremal black hole. $h(\lambda)$ can be Taylor expanded to second-order as
    \begin{eqnarray}
 	    h(\lambda)=\mu_0^2\alpha^2+(2\mu_{0}\delta\mu-8\nu_{0}\delta\nu)\lambda+\left[\mu_{0}\delta^2\mu+(\delta\mu)^2-4\nu_{0}\delta^2\nu-4(\delta\nu)^2\right]\lambda^2+\mathcal{O}(\lambda^3)\label{64}
    \end{eqnarray}
    With the optimal choice of first-order perturbation, we have
    \begin{eqnarray}
 	   0=\delta M-\Phi_{H}\delta Q=\frac{(n-2)\Omega}{16\pi}[\delta\mu-\frac{4\nu_{0}}{\mu_{0}(1+\alpha)}\delta\nu]=\frac{(n-2)\Omega}{16\pi}(\delta\mu-2\sqrt{\frac{1-\alpha}{1+\alpha}}\delta\nu),\label{65}
    \end{eqnarray}
    then the linear term of $\lambda$ in \eqref{64} reads
    \begin{eqnarray}
 	    \left(2\mu_{0}\delta\mu-8\nu_{0}\delta\nu\right)\lambda=(4\sqrt{\frac{1-\alpha}{1+\alpha}}\mu_{0}\delta\nu-4\sqrt{1-\alpha^2}\mu_{0}\delta\nu)\lambda=-4\mu_{0}\delta\nu\alpha\lambda+\mathcal{O}(\alpha^2\lambda).\label{66}
    \end{eqnarray}
    In addition, the left side of inequality \eqref{62} becomes
    \begin{eqnarray}
 	    \delta^2 M-\Phi_{H}\delta^2Q=\frac{(n-2)\Omega}{16\pi}(\delta^2\mu-2\sqrt{\frac{1-\alpha}{1+\alpha}}\delta^2\nu)=\frac{(n-2)\Omega}{16\pi}\left(\delta^2\mu-2\delta^2\nu\right)+\mathcal{O}(\alpha).\label{67}
    \end{eqnarray}
    Meanwhile, note that $\delta'^2M=\delta'^2Q=0$, $\delta'M=\delta M$ and $\delta'Q=\delta Q$ give out
    \begin{eqnarray}
 	    \delta'^2\mu=\delta'^2\nu=0,\quad\delta'\mu=\delta\mu,\quad\delta'\nu=\delta\nu,\label{69}
    \end{eqnarray}
    so the second-order variation of $A^{ST}_B$ can be calculated out as
    \begin{eqnarray}
 	    \delta'^2A^{ST}_{B}&=&\left[(\partial_{\mu}^2A_{B}^{ST})(\delta'\mu)^2+(\partial_{\nu}^2A_{B}^{ST})(\delta'\nu)^2+2(\partial_{\nu}\partial_{\mu}A_{B}^{ST})\delta'\mu\delta'\nu+(\partial_{\mu}A_{B}^{ST})\delta'^2\mu+(\partial_{\nu}A_{B}^{ST})\delta'^2\nu\right]\big|_{\lambda'=0}\nonumber\\
 	    &=&\left[(\partial_{\mu}^2A_{B}^{ST})(\delta\mu)^2+(\partial_{\nu}^2A_{B}^{ST})(\delta\nu)^2+2(\partial_{\nu}\partial_{\mu}A_{B}^{ST})\delta\mu\delta\nu\right]\big|_{\lambda'=0}\nonumber\\
 	    &=&-\frac{n-2}{(n-3)^2}2^{-\frac{n-2}{n-3}}\mu_{0}^{-\frac{n-4}{n-3}}\Omega(1+\alpha)^{\frac{1}{n-3}}\alpha^{-3}\left[\beta_{1}(\delta\mu)^2+\beta_{2}\delta\mu\delta\nu
 	    +\beta_{3}(\delta\nu)^2\right]\nonumber\\
 	    &=&-\frac{n-2}{(n-3)^2}2^{-\frac{n-2}{n-3}}\mu_{0}^{-\frac{n-4}{n-3}}\Omega(\delta\nu)^2(1+\alpha)^{\frac{1}{n-3}}\alpha^{-3}(4\frac{1-\alpha}{1+\alpha}\beta_{1}+2\sqrt{\frac{1-\alpha}{1+\alpha}}\beta_{2}+\beta_{3})\nonumber\\
 	    &=&-\frac{4(n-2)}{n-3}2^{-\frac{n-2}{n-3}} \mu_{0}^{-\frac{n-4}{n-3}}\Omega(\delta\nu)^2\alpha^{-1}+\mathcal{O}(1),\label{70}
    \end{eqnarray}
    where
    \begin{eqnarray}
    	\beta_{1}=(n-3-n\alpha+2\alpha)(1+\alpha),\quad\beta_{2}=-4(n-3-\alpha)\sqrt{1-\alpha^2},\quad\beta_{3}=4(n-3-\alpha+\alpha^2).
    \end{eqnarray}
    On the other hand, the surface gravity  \eqref{34} reads
    \begin{eqnarray}
 	    \kappa
    	&=&\frac{n-3}{2}\Big[\frac{\mu_{0}}{\left(\frac{\mu_{0}}{2}(1+\alpha)\right)^{\frac{n-2}{n-3}}}-\frac{(1-\alpha^2)\mu_{0}^2}{2\left(\frac{\mu_{0}}{2}(1+\alpha)\right)^{\frac{2n-5}{n-3}}}\Big]\nonumber\\
 	    &=&(n-3)2^{\frac{1}{n-3}}\mu_{0}^{-\frac{1}{n-3}}\alpha+\mathcal{O}(\alpha^2).\label{71}
    \end{eqnarray}
     Therefore the second-order inequality \eqref{62} gives rise to 
     \begin{eqnarray}
     	\delta^2\mu-2\delta^2\nu\ge-\frac{2\kappa}{(n-2)\Omega}\delta'^2 A_{B}^{ST}+\mathcal{O}(\alpha)= 4\mu_{0}^{-1}(\delta\nu)^2+\mathcal{O}(\alpha),\label{72}
     \end{eqnarray}
     which will impose a restriction on the quadratic term of $\lambda$ in \eqref{64},
     \begin{eqnarray}
     	&&\left[\mu_{0}\delta^2\mu+(\delta\mu)^2-4\nu_{0}\delta^2\nu-4(\delta\nu)^2\right]\lambda^2\nonumber\\
     	&\ge&\big[\mu_{0}\left(4\mu_{0}^{-1}(\delta\nu)^2+2\delta^2\nu+\mathcal{O}(\alpha)\right)-2\sqrt{1-\alpha^2}\mu_{0}\delta^2\nu+4(\frac{1-\alpha}{1+\alpha}-1)(\delta\nu)^2\big]\lambda^2\nonumber\\
     	&=& 4(\delta\nu)^2\lambda^2+\mathcal{O}(\alpha\lambda^2).\label{73}
     \end{eqnarray}
    Collecting all the above results together, we have
     \begin{eqnarray}
     	h(\lambda)&\ge& \mu_{0}^2\alpha^2-4\mu_{0}\delta\nu\alpha\lambda+4(\delta\nu)^2\lambda^2+\mathcal{O}(\lambda^3,\alpha\lambda^2,\alpha^2\lambda)\nonumber\\
     &=&(\mu_{0}\alpha-2\delta\nu\lambda)^2+\mathcal{O}(\lambda^3,\alpha\lambda^2,\alpha^2\lambda) \ge 0.\label{74}
     \end{eqnarray}
    Thus as we can see, at the linear level of $\lambda$, it is possible to have $h(\lambda)<0$ such that our higher-dimensional charged black hole can be over-charged into naked singularity, in good agreement with the result obtained in \cite{Karl}. However, when the quadratic order correction is taken into account, our higher-dimensional charged black hole cannot be over-charged, invalidating the Hubeny type violation of WCC.
\section{Conclusion}
It is shown in \cite{Karl} that the old version of gedanken experiment can not destroy a higher-dimensional extremal charged black hole but can destroy a higher-dimensional nearly extremal charged black hole. However, in this paper we have invalidated this Hubeny type violation of WCC by appealing to the second order perturbation inequality for the most dangerous first order perturbation in the new version of gedanken experiment, thus WCC is restored in this scenario.

  Together with the optimal result obtained in \cite{Wald17,jincheng}, our result may indicate that no violation of WCC can ever occur at second order when one tries to perform the similar gedanken experiment on any asymptotically flat black hole. Of course, this does not mean that black hole, once formed, can never be destroyed. For example, recent fully non-linear numerical simulation shows that a $6$-dimensional Myers-Perry black hole can pinch off, leading to the formation of naked singularity\cite{NR3}. If our suspicion is correct, then such a violation of WCC must occur at higher orders. But nevertheless, it is apparently valuable to check this by applying the new version of gedanken experiment explicitly to these situations in the future work. 

\section*{Acknowledgments}
This work is partially supported by the National Natural Science Foundation of China with Grant Nos.11675015 and 11775022, as well as by ``the Fundamental Research Funds for the Central Universities" with Grant No.2015NT16. Y. Mo is supported in part by the Beijing Research Fund for Talented Undergraduates. In addition, we would like to thank
Jincheng An, Jieru Shan and Hongbao Zhang for their helpful suggestions and discussions.

\end{document}